\documentclass[3p,preprint, twocolumn]{elsarticle}
\usepackage{amsmath}
\usepackage{graphicx}% Include figure files
\usepackage{makecell}
\usepackage{adjustbox}
\usepackage{epstopdf}
\usepackage{epsfig}
\usepackage{tabu}
\usepackage{dcolumn}% Align table columns on decimal point
\usepackage{bm}% bold math
\usepackage{color}
\usepackage{subfigure}
\usepackage{multirow}
\usepackage{wrapfig}
\usepackage{rotate,color,url}
\usepackage{longtable}
\usepackage{time}
\usepackage{ulem}
\PassOptionsToPackage{hyphens}{url}\usepackage[pagebackref=false,colorlinks,linkcolor=blue,citecolor=blue,urlcolor=blue]{hyperref}
\usepackage{multirow,varwidth}
\usepackage[notransparent]{svg}
\usepackage{subfigure}
\usepackage[T1]{fontenc}
\usepackage{url}

\newcommand{\onlinecite}[1]{\hspace{-1 ex} \nocite{#1}\citenum{#1}}
\makeatletter
\g@addto@macro{\UrlBreaks}{\UrlOrds}
\g@addto@macro{\UrlBreaks}{%
	\do\/\do\d%
}
\makeatother

\DeclareUnicodeCharacter{0308}{HERE!HERE!} 

\begin{document}
\title{Discovery of Novel Silicon Allotropes with Optimized Band Gaps to  Enhance Solar Cell Efficiency through Evolutionary Algorithms and Machine Learning}
	
\author[a]{Mostafa Yaghoobi}
\author[a,b]{Mojtaba Alaei\corref{author}}
\author[a]{Mahtab Shirazi}
\author[b]{Nafise Rezaei}
\author[c,d]{Stefano de Gironcoli}
	
\cortext[author] {Corresponding author.\\\textit{E-mail address:}  m.alaei@iut.ac.ir}
\address[a]{Department of Physics, Isfahan University of Technology, Isfahan 84156-83111, Iran}
\address[b]{Skolkovo Institute of Science and Technology, 121205, Bolshoy Boulevard 30, bld. 1, Moscow, Russia}
\address[c]{Scuola Internazionale Superiore di Studi Avanzati, Trieste, Italy}
\address[d]{CNR-IOM DEMOCRITOS, Istituto Officina dei Materiali, Trieste, Italy}
%\address{Department of Physics, Isfahan University of Technology, Isfahan 84156-83111, Iran}
	
\begin{abstract}
In the pursuit of advancing solar energy technologies, this study presents 20 direct and quasi-direct band gap silicon crystalline semiconductors that satisfy the Shockley-Queisser limit, a benchmark for solar cell efficiency. Employing two evolutionary algorithm-based searches, we optimize structures and calculate fitness function using the DFTB method and Gaussian approximation potential. Following the preselection of structures based on energy considerations, we further optimize them using PBEsol DFT. Subsequently, we screen the structures based on their band gap, employing a DFTB method tailored for band gap calculation of silicon crystals. To ensure accurate band gap determination, we employ HSE and GW methods. To validate the structural stability, we employ phonon analysis via linear regression algorithm applied to PBEsol DFT data. Significantly, the structures unveiled in this study are of great importance due to their proven stability from both mechanical and dynamic perspectives. Furthermore, the ductility and low density of certain structures enhance their potential application. We examine the optical properties by studying the imaginary part of the dielectric function by solving the Bethe-Salpeter Equation on top of GW approximation. By calculating the SLME, we achieve an efficiency of 32.7\% for Si$_{22}$ at a thickness of 500 nm.  Moreover, the study harnesses various machine learning algorithms to develop a predictive model  for the band gap energy of these silicon structures. Input data for machine learning models are derived from structural MBTR and SOAP descriptors, as well as DFT outputs. Notably, the results reveal that features extracted from DFT outperform the MBTR and SOAP descriptors.
\end{abstract}
	
\maketitle
\section{Introduction}
In 2016, the Paris Agreement established a goal of limiting the global average temperature increase to well below $2^\circ$C above pre-industrial levels, 
with a further aim of restricting it to $1.5^\circ$C. 
Fulfilling this crucial mission requires the development of alternative, clean energy sources that can reduce our dependence on fossil fuels. 
Solar energy is a prime candidate for a sustainable and eco-friendly energy source, and photovoltaic (PV) cells are key to converting solar energy into electricity.  
However, the efficiency of PV cells is limited by the properties of the materials used to construct them, particularly their band gap. %The efficiency of solar cells directly depends on the band gap energy of the materials that make up the cell.
Moreover, employing direct and quasi-direct bandgap semiconductors in solar cells results in a diminished thickness of the solar cell. 
In these materials, there is enhanced absorption of incident photons, leading to increased cell efficiency~\cite{SREEDEVI2019350}. 
According to the Shockley–Queisser limit~\cite{RUHLE2016139},  solar cell efficiency surpasses 30 percent when the band gap of the material used in the cell falls within the range of 0.93 to 1.61eV, with the highest efficiency occurring at the band gap of 1.34 eV.  
While silicon has historically been the primary material for PV cells due to its abundance and stability, 
its indirect band gap energy of 1.1 eV poses limitations on its efficiency. 
Therefore, finding silicon structures that have a direct and quasi-direct band gap energy and are within the Shockley–Queisser limit will be an important step in the development of solar cells and in increasing their efficiency.
	
Predicting crystal structures has always been a significant challenge in material science. 
In Ref.~\onlinecite{C6CP00195E}, Fan et al. proposed two phases of metastable silicon structures by substituting the Si atom in the carbon structures. 
Zhang {\sl et al.}~\cite{ZHANG2020103271} proposed four semiconducting silicon allotropes 
with direct and quasi-direct band gaps between 1.193 and 1.473 eV, characterized by low energy. 
Chun-Xiang Zhao {\sl et al.}~\cite{ZHAO2019125903} have introduced a new silicon allotrope with a direct band gap of 0.591eV. 
Meanwhile, Wang {\sl et al.}~\cite{WangT36silicon} suggested six metastable silicon structures using 
the particle swarm optimization algorithm. Additionally, using the inverse band structure design approach based on 
the particle swarm optimization algorithm, H. J. Xiang {\sl et al.} were able to predict a 20-atom silicon phase 
with a quasi-direct gap of 1.55 eV~\cite{PhysRevLett.110.118702}. Moreover, in a study by Q. Wei {\sl et al.}~\cite{C9CP03128F}, 
researchers explored crystalline silicon structures with direct gaps by replacing carbon atoms with silicon in the existing carbon phases, 
yielding six silicon allotropes with direct band gaps. Also, in Refs.~\onlinecite{PhysRevB.92.014101} 
and \onlinecite{PhysRevB.86.121204}, researchers introduced several metastable silicon crystal structures with direct and quasi-direct band gaps within the range of 1 to 1.8 eV, employing the ab initio minima hopping method.
%\textcolor{red}{Please add the ref. PhysRevB.86.121204 and PhysRevB.92.014101. We have them inside refrence.bib but we do not mention them here. 
%I think we definitely should mention these studies.} 

In this study, we used the power of the evolutionary algorithm (EA)~\cite{LONIE2011372,BAHMANN20131618,10.1063/1.5037159} 
to predict novel phases of silicon crystal structures, a method that has recently captured the attention of researchers.  
Through two distinct EA-based searches, we explored silicon structures ranging in size from 2 to 45 atoms.  
In the first search, we employed the self-consistent-charge density-functional tight-binding 
(DFTB)~\cite{DFTB_an_approximate_DFT,PhysRevB.58.7260} method to calculate the fitness function, 
which in this study is the enthalpy of the structures. This choice aimed to ensure the stability of the predicted structures. 
In the second search, the Gaussian approximation potential (GAP)~\cite{PhysRevLett.104.136403} was used to calculate the fitness function. 
The GAP potential significantly accelerated the enthalpy predictions, thereby facilitating the exploration of larger structures.
	
Subsequently, our examination focused on identifying silicon structures that align with the Shockley–Queisser limit, thereby demonstrating the potential for enhanced solar cell efficiency. Through a comprehensive analysis of the dynamical and mechanical stability of these structures, we sought to guarantee their practicality and reliability. Ultimately, we meticulously screened and identified a collection of 20 unique and promising silicon structures, each holding the potential to reshape the landscape of solar cell technology.
	
The paper is structured as follows: we begin by detailing our methodology,  casting the foundation for the subsequent results. We then unveil the silicon structures we have discovered and describe their mechanical and electronic properties. Subsequently, our discussion addresses the implications of these findings and highlights the most promising structures for solar cell applications by examining the optical and electrical properties. Finally, we delve into the prediction of band gap energy for these structures, utilizing various machine learning(ML) models. %Our focus remained on structures featuring a direct band gap within the critical range of 0.93 to 1.61 eV, a crucial criterion for advancing the development of efficient solar cells.  

\section{Computational Details}

In the pursuit of crystal structure prediction, we harnessed the well-established USPEX code~\cite{GLASS2006713}, 
which has demonstrated promising results and is extensively employed in materials science researches~\cite{OGANOV200695,10.2138/rmg.2010.71.13}.

As mentioned, Our study encompassed two distinct searches, each wielding distinct computational approaches. In the first search, 
to calculate the enthalpy of the structures we used DFTB with pbc Slater-Koster~\cite{Sieck_pbc_SK} (SK), 
executed through the DFTB+ software package~\cite{AradiDFTB+}, 
which offers a computationally efficient alternative to first-principles DFT methods. 	
In the subsequent search, we replaced DFTB with GAP~\cite{PhysRevX.8.041048}. 
GAP's formalism rests on the Gaussian process regression (GPR) and smooth overlap atomic position (SOAP)~\cite{PhysRevB.87.184115} kernel.  
To implement the GAP and incorporate it with USPEX, we used the LAMMPS code~\cite{PhysRevLett.104.136403,Bartk-Prtay2010}.

%At each stage of the EA, local optimization procedures were executed to eliminate energetically unfavorable structures from the evolving population. 
%The relaxation procedure was initially executed employing the Broyden-Fletcher-Goldfarb-Shanno (BFGS)\cite{pfrommer1997relaxation} minimization scheme during the first search. We established a force convergence threshold of 0.0008 atomic units (a.u) and employed pbc Slater-Koster\cite{Sieck_pbc_SK}, which are suitable for crystal structure electronic calculation. During the second search, the conjugate gradient (CG)\cite{PhysRevB.39.4997} minimization scheme was utilized to attain energy optimization and structure relaxation.
	
In the first search, we set the initial population size to $250$, the population of each generation to $200$ structures, 
the number of generations to $100$, and a stopping criterion, whereby the algorithm would terminate 
if there was no improvement in the best structures after 30 generations. 
Furthermore, in our pursuit of more optimized structures, 
we employed the USPEX once more to predict atomic structures comprising 2 to 20 atoms. 
In this time, we employed the structures procured during the preceding search as initial seeds within the genetic algorithm framework. 
We carried out two generations of the genetic algorithm, with a population size of $50$ in each generation. 
For fitness function calculations and structural relaxation, we employed DFT via Quantum Espresso (QE)~\cite{Giannozzi_2009}.
	
In the second search, we set the initial population size to 300 and the population of each generation to 250 structures. 
We specified the number of generations to 200, and a stopping criterion, whereby the algorithm would terminate if there was no improvement in the best structures after 30 generations. Overall, according to the enthalpy and volume of the structures, 
the first search led to the production of 536 structures, and the second search led to the production of 827 structures. 
	
%In the initial search, 40\% of structures in each generational iteration were generated via heredity, 30\% through random symmetry operations, and the remaining 30\% through softmutation. In the subsequent search, we modified the distribution to enhance the contribution of heredity, with 55\% of structures in each generation derived through this mechanism, 22\% through random symmetry operations, and 23\% via softmutation. 
%Since the structures were only made of the same type of atom (Si), it is not necessary to use permutation. Moreover, we implemented a stopping criterion, whereby the algorithm would terminate if there was no improvement in the best structures after 30 generations.
	
To ensure that equilibrium positions were reached for each structure, all structures (1263 Si structure) were relaxed by QE using projector augmented wave method (PAW) with higher accuracy. 
We set  PBEsol functional for exchange-correlation appximation, a uniform reciprocal space mesh point spacing of 0.15 (1/\AA),  40 Ry  kinetic energy cutoff  for wavefunctions, and 400 Ry kinetic energy cutoff for charge density.
%During the final relaxation stage, the BFGS minimization algorithm was employed. Our selection of parameters for this DFT-based structural relaxation included a force convergence threshold of $1.0\times 10^{-4}$ atomic units (a.u), a kinetic energy cutoff of 40Ry for wavefunctions, a kinetic energy cutoff for charge density of 400Ry.  Notably, uniform values for these parameters were consistently applied across all DFT calculations, employing the PBEsol functional. Moreover, the selection of a uniform reciprocal space mesh point spacing of 0.15 (1/\AA) was a common practice adopted throughout this study.
	
After relaxing the structures, we investigated the band gap energy of the structures to find suitable structures for solar cells. 
Despite the satisfactory accuracy of DFT in material energy calculations, this method leads to underestimation in band gap energy prediction. 
In order to overcome the band gap problem, methods such as GW~\cite{PhysRev.139.A796} and 
the Heyd-Scuseria-Ernzerhof (HSE) hybrid functional~\cite{10.1063/1.464304} have been suggested, 
although they are computationally demanding. For instance, in the studies by R.I. Eglitis et al.\cite{ma16247623,EGLITIS2018459}, the researchers employed two hybrid exchange functionals, along with Gaussian-type localized basis sets, to determine the band gaps of various perovskites. Their findings closely align with the data obtained from experimental observations.
More recently, researchers have turned to the DFTB method, leveraging the Hamiltonian and overlap matrix parameters of atomic orbitals, which are pre-tabulated and calculated using empirical methods. This approach offers a faster and more cost-effective means to obtain the band structure of materials. 
In our study, we utilized the DFTB method, employing si-band Slater-Koster~\cite{7014245} fitted for band structure calculations, to compute the band structure of the crystal structures. 
To assess the accuracy of the DFTB method in comparison to the GW and HSE hybrid functional methods, 
we selected four structures previously identified as promising candidates for use in solar cells and computed their band structure 
within the self-consistent GW using the Yambo code~\cite{MARINI20091392}. 
The resulting band gap energy values for each structure and each method (HSE, GW and DFTB) are presented in the Table~\ref{hsegwdftb}. 
\begin{table}[h]\tiny
	\centering
	\caption{\small Values of the band gaps in eV that obtaind by HSE, GW and DFTB. The structures are extracted from Ref \cite{10.1021/ja5035792}.}
	\resizebox{\columnwidth}{!}{%
		\begin{tabular}{ |c|c|c|c|  }
			\hline
			\hline
			structure&HSE\cite{10.1021/ja5035792}&GW&DFTB\\
			\hline
			mC12& 1.24 &1.33&1.31\\
			oC12& 0.96&1.03&1.19  \\
			tl16& 1.25&	1.33&1.09  \\
			tp16&0.91 &0.90&1.19\\
			\hline
	\end{tabular}}\label{hsegwdftb}
\end{table}
The table clearly illustrates that the DFTB method based on si-band SK exhibits a good level of accuracy when compared to the other two methods. The mean squared error between the HSE values and DFTB is 0.04 eV, and it is 0.042 eV between the DFTB and GW values.
Importantly, DFTB stands out for its significantly reduced computational time requirements. 
It is worth mentioning that we use Wannier90~\cite{Wannier90} to obtain the HSE band structures from HSE scf calculations of QE.

In our research, we also conducted a comprehensive assessment of the mechanical and dynamic stability of the screened crystal structures. 
To achieve this, we employed DFT calculations using the THERMO-PW package code~\cite{thermopw}. 
The DFT calculations were performed, using the PBEsol pseudopotential~\cite{prandini2018precision}. 
Dynamically stable structures can withstand small atomic vibrations, a property we examined by studying their phonon properties. 
We utilized the ML-based hiphive code~\cite{https://doi.org/10.1002/adts.201800184} to check the phononic properties of the structures.

In addition, the optical properties of the structures have been studied by investigating their dielectric function by solving the Bethe-Salpeter equation\cite{PhysRev.84.1232,Sangalli_2019} on top of previous GW calculations using Yambo. We set the energy cutoff for the calculation of the static dielectric function to 8 Ry. In the GW calculations, we set the distance between k points in the reciprocal space to 0.15 (1/Å) for all structures. This resulted in a 14×14×14 mesh grid for the Si$_{\text{diamond}}$. However, due to the computational intensity, we used a 12×12×12 mesh grid for the Si$_{\text{diamond}}$ in this specific case, but the mesh grid for the rest of the structures remained unchanged. 

Furthermore, We conducted a study aimed at developing an ML model to predict the band gap energy of silicon crystal structures. 
In our investigation, we explored various ML algorithms to determine their effectiveness in predicting the band gap of these structures. 
The algorithms considered in our analysis included Multilayer Perceptron Neural Network that 
we simply denote by NN, Kernel Ridge Regression (KRR), Support Vector Regression (SVR), Convolutional Neural Networks (CNN), 
as well as Decision Tree (DT), Random Forest (RF), and XGbosst. 
We implemented NN and CNN using the TensorFlow framework~\cite{tensorflow_developers_2023_8306789}, while for SVR, DT, RF, and XGBoost, 
we utilized the Scikit-Learn library~\cite{scikit-learn} in Python.
In this study, we use two important structural descriptors, Many-body tensor representation(MBTR)~\cite{Huo_2022} 
and SOAP~\cite{descriptorslocalglobal2018}, to construct suitable features for ML models. 
To generate these descriptors for the crystal structures, we utilized the Dscribe computational package~\cite{HIMANEN2020106949}

\section{Results}
\subsection{Structure Screening}
\begin{figure}[htbp]
	\centerline{\includegraphics[scale=.3]{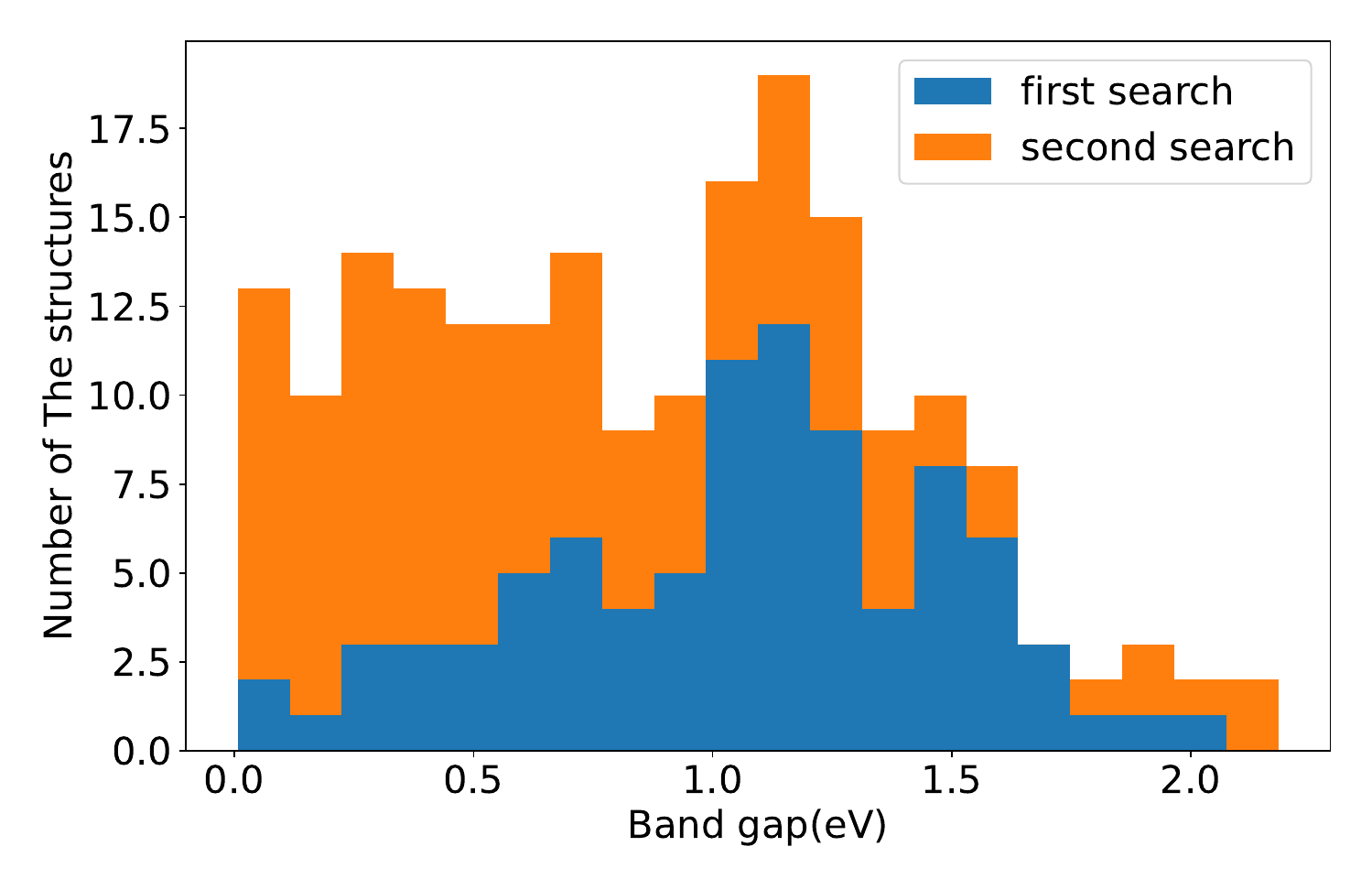}}
	\caption{The histogram of the band gap energies for the structures in both searches.}
	\label{band_gap_hist}
\end{figure}

In our initial search, we identified 148 semiconductor structures featuring band gaps ranging from 0.0949 to 2.014 eV.
In the second search, it turns out that there are 256 semiconductor structures with band gap between 0.007 and 2.183eV. 
To ensure the uniqueness and distinctiveness of our selection, we conducted a thorough examination of the structural geometry. 
This scrutiny led to the elimination of similar structures and those that proved identical to the Si$_{\text{diamond}}$. 
This process was done by comparing the Valle-Oganov fingerprint~\cite{valle2010crystal} of each structure with other structures 
and theSi$_{\text{diamond}}$. 
Therefore, the number of unique semiconductor structures was reduced to $88$ in the first search and to 108 structures in the second search. %The band gap energy values for the structures in both searches are illustrated in Fig\ref{band_gap_energies}.
%This distinction is evident in 
The band gap histogram of the structures for both searches depicted in Figure~\ref{band_gap_hist}.
The average band gap energy value in the initial search equals 1.06 eV, while in the subsequent search, it is 0.72 eV. 
%%%%%%%Fig 1
%%%%%%
Further narrowing our focus, we identified 85 structures that have band gaps between 0.93 and 1.61 eV. 
Screening these structures for those whose band gap is direct and quasi-direct, we found a subset of 20 structures suitable for further consideration.

\begin{figure*}[h]
	\centering%
    \includegraphics[scale=.2]{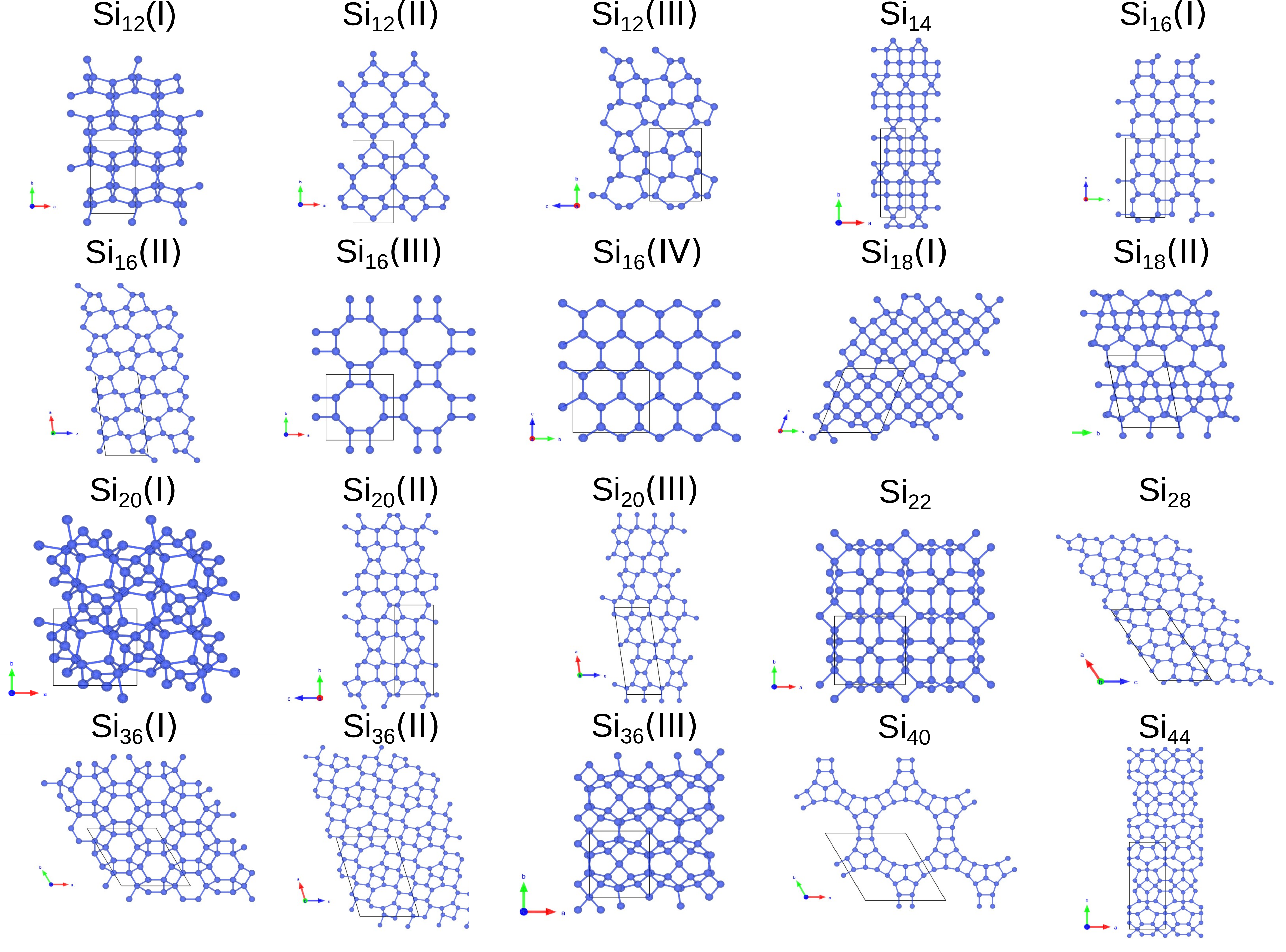}
	\caption{The ball-and-stick representations of the candidate crystal structures.
		 The space group of these structures is presented in Table~\ref{mechtab}.}
	\label{structures_visual}
\end{figure*}

In addition, we conducted a geometric analysis of all 20 screened final structures through $\texttt{Spglib}$\cite{togo2018spglib}. By increasing the tolerance for the Cartesian coordinate distances between atoms, we achieved geometric refinement, resulting in symmetric structures. Following this refinement, we performed another optimization of the structures using PBEsol DFT.
The ball-and-stick drawings of the crystal structures are presented in Figure~\ref{structures_visual}. 

Furthermore, an examination of the fingerprint of all 20 structures was conducted in comparison to the pure silicon structures introduced in the Material Project database. 
This analysis revealed that Si$_{16}$(II) had previously been documented on the Material Project platform. 
Moreover, structures Si$_{12}$(II) and Si$_{16}$(III) exhibited a noteworthy similarity to structures oC12 and tl16, respectively, 
as previously elucidated in Ref.~\onlinecite{10.1021/ja5035792}.

It is worth highlighting that among these structures, Si$_{40}$, Si$_{12}$(II), Si$_{16}$(III), Si$_{20}$(III), Si$_{16}$(II), Si$_{16}$(I), Si$_{36}$(II), and Si$_{36}$(I) hold particular significance due to their classification as porous crystals. 
These structures exhibit void tunnels, a distinct feature in their composition. The void tunnels in Si$_{40}$, Si$_{12}$(II), Si$_{36}$(II), Si$_{16}$(I), Si$_{16}$(II), and Si$_{20}$(III) structures are formed by connecting 4, 5, and 6-membered rings, and are observable from one side. However, in the case of Si$_{36}$(I), the void tunnels exist on two sides, and they are formed by connecting 4-membered rings.

Direct and quasi-direct semiconductors are particularly advantageous in the fabrication of thin-film solar cells, 
obviating the need for a thick layer required in the use of indirect band gap semiconductors~\cite{SREEDEVI2019350}. 
To evaluate the band gap properties of various semiconductor structures, we conducted an in-depth analysis of their band structures. 
The HSE and GW band structures can be found in Supplementary Information (SI).
The results of this analysis are summarized in Table~\ref{bandgaps}, 
wherein DFTB, HSE, and GW results were showed for a comprehensive examination.
\begin{table*}[h]
	\centering
	\caption{\small The band gap energies of the structures determined via DFTB, HSE, and GW calculations. Additionally, the band gap energies of the structures predicted using ML models including CNN, XGBoost, and RF, in combination with DFT outputs features. In the second column, we indicate the structure type by specifying whether it is a direct(D) or quasi-direct(QD) semiconductor.}
	%\resizebox{\textwidth}{!}{
	%\tiny
	\small
	\begin{tabular}{|l|*{7}{c|}}
		\hline
		\multirow{2}{*}{structures}&\multirow{2}{*}{type}&\multicolumn{3}{c|}{band gap}&\multicolumn{3}{c|}{ML band gap} \\
		\cline{3-8}
		&& DFTB & HSE& GW & CNN&XGBoost&RF\\
		\hline
		Si$_{12}$(I)&QD&1.28&1.46&1.08&1.08&1.27&1.17\\
		Si$_{12}$(II)&D&1.24&1.16&0.75&1.24&1.21&1.15\\
		Si$_{12}$(III)&QD&1.49&1.63&1.34&1.56&1.49&1.47\\
		Si$_{14}$&D&1.13&1.06&0.85&0.91&1.11&1.01\\
		Si$_{16}$(I)&D&0.97&1.15&0.93&1.11&0.90&0.90\\
		Si$_{16}$(II)&QD&0.95&0.92&0.73&0.72&0.68&0.59\\
		Si$_{16}$(III)&D&1.16&1.35&1.19&1.24&1.11&1.17\\
		Si$_{16}$(IV)&D&1.18&1.31&1.11&1.11&1.07&1.08\\
		Si$_{18}$(I)&QD&1.43&1.6&1.42&1.37&1.27&1.34\\
		Si$_{18}$(II)&D&1.03&1.46&1.30&1.24&1.05&1.20\\
		Si$_{20}$(I)&D&1.22&1.31&1.03&1.27&1.21&1.21\\
		Si$_{20}$(II)&D&1.26&1.45&1.20&1.31&1.25&1.23\\
		Si$_{20}$(III)&QD&1.16&1.31&1.06&1.20&1.21&1.16\\
		Si$_{22}$&D&1.4&1.57&1.34&1.49&1.35&1.24\\
		Si$_{28}$&QD&1.38&1.50&1.23&1.31&1.36&1.29\\
		Si$_{36}$(I)&QD&1.15&1.43&1.28&1.47&1.25&1.27\\
		Si$_{36}$(II)&D&1.14&1.28&1.17&1.10&1.05&1.08\\
		Si$_{36}$(III)&D&1.13&1.34&1.20&1.17&1.12&1.14\\
		Si$_{40}$&QD&1.34&1.35&1.35&1.22&1.20&1.24\\
		Si$_{44}$&QD&1.38&1.24&1.07&1.32&1.36&1.28\\
		Si$_{\text{diamond}}$&ID&1.14&1.28&1.10&1.11&1.26&1.23\\
		\hline
	\end{tabular}%}
	\label{bandgaps}
\end{table*}

%new
The DFTB has a 14\% error, and the HSE has a 20\% error when compared to GW results. 
This suggests that the DFTB method is reliable for estimating the band gap.
%new
In addition, Table~\ref{bandgaps} contains the values of band gap energy predicted by three of the best ML models, including (CNN, XGBoost, and RF)+DFT output. 
According to our findings, 11 structures including Si$_{16}$(III), Si$_{12}$(II), Si$_{16}$(I), Si$_{22}$, Si$_{36}$(II), Si$_{14}$, Si$_{18}$(II), Si$_{36}$(III), Si$_{20}$(II), Si$_{20}$(I), and Si$_{16}$(IV), exhibit direct band gaps, while the remaining structures fall into the category of quasi-direct semiconductors. 
To ascertain whether a structure qualifies as a quasi-direct semiconductor, 
we compute the direct band gap  (E$_g^d$) and the band gap (E$_g$)
If the result of E$_g^d$ - E$_g$ is less than 0.15 eV, 
the system is classified as a quasi-direct band gap semiconductor~\cite{PhysRevB.90.115209}. 
The corresponding values of E$_g$,  E$_g^d$, and E$_g^d$ - E$_g$ 
for the quasi-direct  semiconductors obtained through HSE method are presented in the Table~\ref{quasi_directs}.
\begin{table}[h!]
	\centering
	\caption{\small Band gap energy (E$_g$) and direct band gap (E$_g^d$) values of the structures that were known quasi-direct semiconductor based on HSE calculations.}
	%\resizebox{\textwidth}{!}{
	\begin{tabular}{|l|c|c|c|}
		\hline
		structure& E$^d_g$ & E$_g$ & E$^d_g$ - E$_g$\\
		\hline
		Si$_{12}$(I)&1.50&1.46&0.043\\
		Si$_{12}$(III)&1.64&1.63&0.008\\
		Si$_{16}$(II)&1.06&0.92&0.139\\
		Si$_{18}$(I)&1.65&1.60&0.044\\
		Si$_{20}$(III)&1.32&1.31&0.005\\
		Si$_{28}$&1.50&1.50&0.003\\
		Si$_{36}$(I)&1.52&1.43&0.094\\
		Si$_{40}$&1.39&1.35&0.043\\
		Si$_{44}$&1.24&1.23&0.006\\
		\hline
	\end{tabular}%}
	\label{quasi_directs}
\end{table}

It is important to emphasize that there are cases where the results of GW and HSE analysis differ.
Specifically, an examination of the GW calculations for the Si$_{16}$(IV) structure reveals it 
to be a quasi-direct band gap with a direct gap of 1.13 eV. 
However, HSE results designate this structure as a direct semiconductor. 
In the case of Si$_{18}$(II), HSE calculations categorize it as a direct semiconductor, 
while GW analysis identifies it as a quasi-direct semiconductor with a direct gap of 1.37 eV. 
A similar pattern is observed in the assessment of Si$_{22}$, 
where the direct gap is determined to be 1.36 eV through GW calculations.
Furthermore, Si$_{28}$ and Si$_{20}$(III) structures are identified as direct semiconductors according to GW, 
but HSE calculations classify them as quasi-direct semiconductors.

\subsection{Structural Properties}
Table \ref{mechtab} presents the data encompassing bulk modulus ($B$), 
shear modulus ($G$), Young modulus ($E$), Poisson ratio ($n$), 
and the ratio of bulk modulus to shear modulus($B$/$G$) in the Voigt-Reuss-Hill average approximation. 
\begin{table*}[h]
	\centering
	\caption{\small The table comprises bulk modulus ($B$), Young modulus ($E$), shear modulus ($G$) measured in gigapascals (GPa), 
		Poisson ratio ($n$), the ratio of bulk to shear modulus ($B$/$G$), density ($\rho$) expressed in 
		grams per cubic centimeter (g/cm$^3$), and relative enthalpy ($\Delta H$) per atom expressed in electronvolts (eV). 
		Last column shows the space group of structures.}
	\resizebox{\textwidth}{!}{
		\small\tiny
		\begin{tabular}{ |l|c|c|c|c|c|c|c|c| }
			\hline
			structure&$B$(GPa)&E(GPa)&G(GPa)&$n$&$B$/$G$&$\rho(g/cm^3)$&$\Delta H\text{(eV per atoms)}$&space group\\
			\hline
			Si$_{12}$(I)&68&82&32&0.297&2.142&2.1428&0.2165&C2/m\\
			Si$_{12}$(II)&83&132&54&0.233&1.54&2.1650&0.0924&C222$_1$\\
			Si$_{12}$(III)&87&134&54&0.243&1.617&2.2866&0.0028&Pmn2$_1$\\
			Si$_{14}$&95&154&63&0.229&1.514&2.5256&0.0104&C222\\
			Si$_{16}$(I)&89&131&52&0.256&1.720&2.2727&0.0644&Pmmm\\
			Si$_{16}$(II)&85&132&53&0.241&1.595&2.2452&0.0710&C2/m\\
			Si$_{16}$(III)&80&108&42&0.272&1.881&2.0726&0.1419&I4/mcm\\
			Si$_{16}$(IV)&85&115&45&0.276&1.900&2.2054&0.0083&Cmcm\\
			Si$_{18}$(I)&87&137&55&0.237&1.571&2.3230&0.0691&P-1\\
			Si$_{18}$(II)&76&112&45&0.254&1.704&2.2050&0.0145&P-1\\
			Si$_{20}$(I)&81&122&49&0.247&1.648& 2.3782&0.0121&P4$_3$2$_1$2\\
			Si$_{20}$(II)&87&134&54&0.244&1.621&2.2285&0.0039&Imm2\\
			Si$_{20}$(III)&85&134&54&0.236&1.562&2.2242&0.0707&C2/m\\
			Si$_{22}$&83&115&45&0.268&1.829&2.1605&0.1382&P4$_2$/mcm\\		
			Si$_{28}$&87&132&53&0.246&1.641&2.2896&0.0045&C2/m\\
			Si$_{36}$(I)&51&51&18&0.334&2.675&1.6026&0.3854&R3m\\	
			Si$_{36}$(II)&83&132&53&0.235&1.560&2.2913&0.0876&C2/m\\
			Si$_{36}$(III)&82&117&46&0.263&1.777&2.1329&0.0091&Ibam\\
			Si$_{40}$&51&62&24&0.296&2.156&1.5358&0.2504&P6$_3$/mcm\\
			Si$_{44}$&65&89&35&0.271&1.859&1.9308&0.0151&Cmcm\\
			Si$_{\text{diamond}}$&91&151&62&0.223&1.472&2.3291&0.0000&Fd-3m\\
			\hline
	\end{tabular}}
	\label{mechtab}
\end{table*}

The bulk modulus reflects the volumetric elasticity of the materials, 
while Young modulus characterizes their tensile elasticity. 
Shear modulus, on the other hand, signifies the material's susceptibility 
to shearing when subjected to opposing forces. 
The ratio of bulk modulus to shear modulus serves as an indicator of the material's brittleness or ductility. 
Specifically, if this ratio falls below 1.75, 
the material is categorized as brittle, whereas values exceeding this threshold denote ductility~\cite{doi:10.1080/14786440808520496}. 
Furthermore, Poisson ratio can also provide insights into the material's brittleness or ductility, with materials exhibiting a Poisson ratio exceeding 0.26 typically classified as ductile~\cite{doi:10.1080/09500830500080474}.

Based on the information presented in Table \ref{mechtab}, an analysis of the $B$/$G$ and $n$ values for structures 
Si$_{36}$(I), Si$_{40}$, Si$_{12}$(I), Si$_{16}$(III), Si$_{22}$, Si$_{36}$(III), Si$_{16}$(IV), and Si$_{44}$ suggests 
that these structures exhibit ductile properties.

Furthermore, when examining the $B$, $E$, and $G$ values, it becomes evident that structures Si$_{36}$(I) and Si$_{40}$ have lower values compared to the other structures. This can be attributed to the porous nature of these two crystal structures.

%The density values provided in Table\ref{mechtab} indicate that Si$_{40}$ and Si$_{36}$(I) possess lower densities in comparison to the other structures. This can be attributed to the presence of tunnel voids within these particular structures.

Moreover, the relative enthalpy values presented in Table~\ref{mechtab}, which are computed using the formula 
$\Delta H = H/N - H_{\text{diamond}}/N_{\text{diamond}}$, where $H$ and $N$ represent the enthalpy 
and the number of atoms in the respective structure, and $H_{\text{diamond}}$ and $N_{\text{diamond}}$ 
correspond to the enthalpy and the number of atoms in the Si$_{\text{diamond}}$, 
offer valuable insights into the relative stability of these crystal structures. 
It is noteworthy that Si$_{12}$(III) emerges as the most stable structure, 
with Si$_{20}$(II) and Si$_{28}$ closely following in terms of stability.

To ensure the stability of the newly developed structures, 
we undertook an investigation focused on analyzing phonon scattering under varying pressures (-5, 0.0,  and 5 GPa). 
Initially, we allowed the structures to relax at two distinct pressures: 5 and -5 GPa. 
Subsequently, we examined the phonon dispersion of the structures for each pressure condition. 
The outcomes revealed that all structures demonstrated stability under the specified pressures, with the exception of Si$_{16}$(III), which exhibited instability at 5 GPa pressure.

\begin{figure*}[h]
	\centering%
	\includegraphics[scale=0.4]{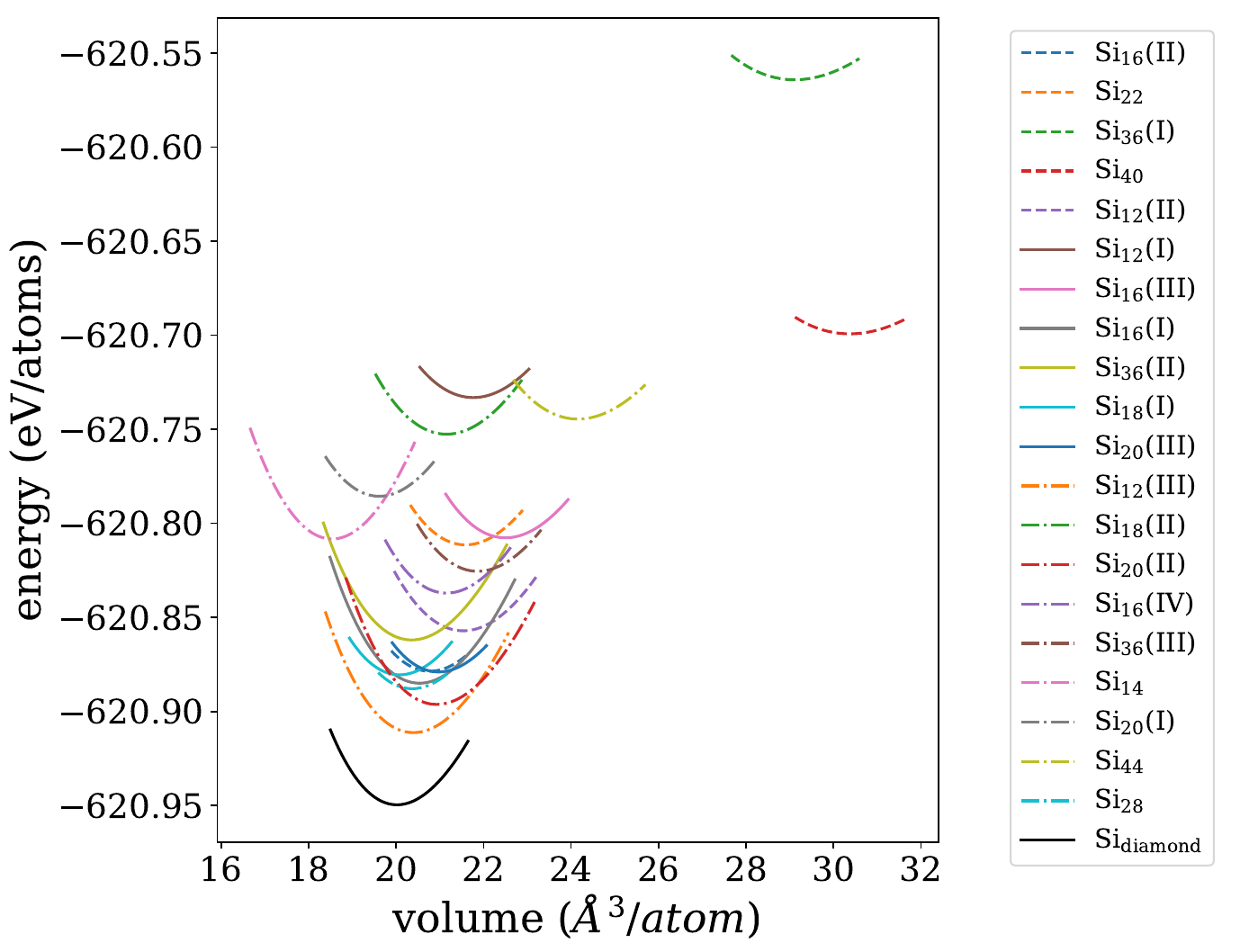}
	\caption{Energy of the structures as a function of volume.}
	\label{energy_volume}
\end{figure*}

Figure~\ref{energy_volume} depicts the energy variations of the structures in relation to volume changes. 
As expected, the Si$_{\text{diamond}}$ demonstrates the most stability when compared to the other structures.
Subsequently, Si$_{12}$(III) demonstrates a higher degree of stability in comparison to the remaining configurations, 
securing the second-highest stability ranking. Following closely behind is Si$_{20}$(II), which attains the next most favorable stability ranking.

In contrast, Si$_{36}$(I) emerges as the least stable configuration, as indicated by the graph's data. Interestingly, Si$_{40}$ exhibits relatively minimal energy fluctuations in response to volume changes. According to Figure~\ref{energy_volume}, it is evident that structures Si$_{40}$ and Si$_{36}$(I) exhibit distinct characteristics compared to other structures. These distinctions primarily arise from their considerable volume, which is indicative of their porous nature. In order to ascertain the stability of these two structures, a thorough investigation of their phonon frequencies across various volumes was conducted.
To achieve this, we systematically generated eight distinct structures for each of the aforementioned Si$_{40}$ and Si$_{36}$(I) structures, each featuring different volumes within the range of 95\% to 114\% of the original structures' volume. Subsequently, we relaxed these structures to determine the equilibrium positions of their constituent atoms under varying volumes and subsequently computed their respective phonon frequencies.
A notable discovery arising from this investigation was the absence of any imaginary frequencies within the calculated phonon spectra for these structures across the entire range of volumes examined. This compelling result strongly supports the stability of both the Si$_{40}$ and Si$_{36}$(I) structures.
\subsection{Optical properties}

\begin{figure}
	\centering%
	\includegraphics[scale=.3]{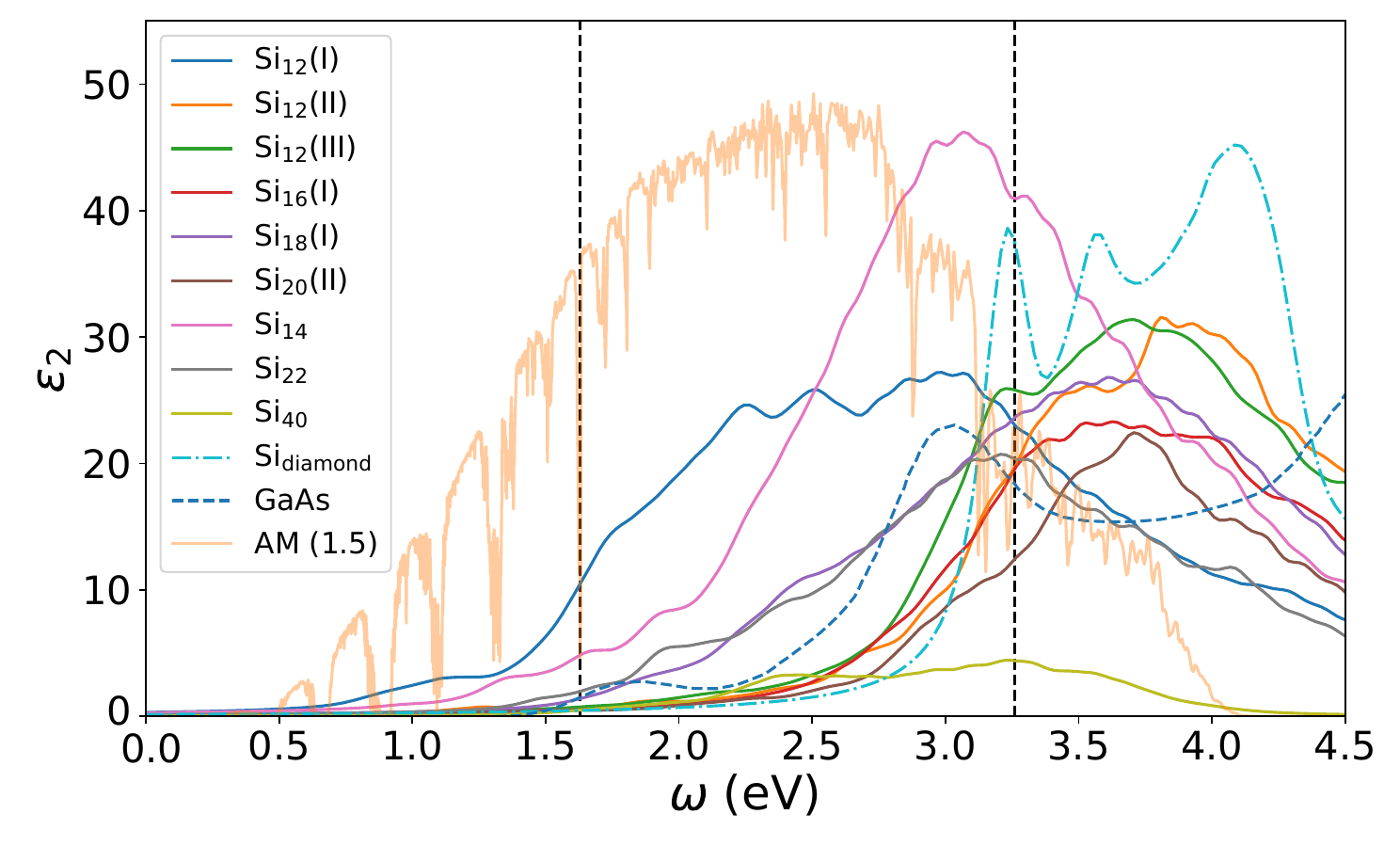}
	\caption{The imaginary part of the dielectric function for promising structures is presented. Additionally, the reference AM 1.5 solar spectral irradiance is plotted in arbitrary units to highlight the region of interest. The two vertical dashed lines correspond to photons with energies of 1.63 eV and 3.2 eV, respectively, from left to right, representing the visible region. The data for GaAs was obtained from Ref.~\cite{PhysRevB.68.205112}. }
	\label{absoption}
\end{figure}

Figure \ref{absoption} shows the imaginary part of the dielectric function for some of the most promising allotropes. For comparison, the diagram also includes the imaginary part of the dielectric function for Si$_{\text{diamond}}$ and GaAs (Gallium arsenide)\cite{PhysRevB.68.205112}, as well as the reference AM 1.5 solar spectral Irradiance. According to the solar spectrum, all structures perform better at absorbing low-energy photons compared to Si${_\text{diamond}}$. 
Additionally, Si${12}$(I), Si${14}$, Si${22}$, and Si$_{18}$(I) show better absorption than GaAs.

To ensure the suitability of the structures to create an single-junction solar cell, it is very important to check their spectroscopic limited maximum efficiency (SLME)\cite{kumarPCL,C6CP03468C}. 
%Because this quantity includes the physical, electrical and optical information of materials, it is of great importance. 
The SLME calculation formalism is based on several important quantities including the fraction of radiative recombination ($f_r$), which is obtained from 
%$f_r = \exp\left(-\frac{E_g^d - E_g}{k_BT}\right)$, 
$f_r = e^{-\frac{\Delta E}{k_BT}}$, 
where $\Delta E=E_g^d-E_g$, $k_B$ and $T$ denote the Boltzmann constant and device temperature, respectively; the absorption spectrum, which is calculated directly from the dielectric function; and the temperature and thickness of the material.
According to $f_r$, when the difference between $E_g^d$ and $E_g$ is large (e.g., in the case of Si$_{\text{diamond}}$, where this value is around 2.3 eV), $f_r$ tends to zero quickly. This leads to a significant increase in reverse saturation current. Therefore, this formalism is not suitable for the structures where the difference between $E_g$ and $E_g^d$ is large. However, it can be used for structures with direct and quasi-direct band gap. 
%In this work, we set the value of $f_r$ to $10^{-3}$, as recommended in Ref. \onlinecite{C6CP03468C} for Si diamond.
To compare our results with silicon diamond, we set the value of $f_r$ to $10^{-3}$, as recommended in Ref. \onlinecite{C6CP03468C} for silicon diamond.

\begin{figure}
	\centering%
	\includegraphics[scale=.3]{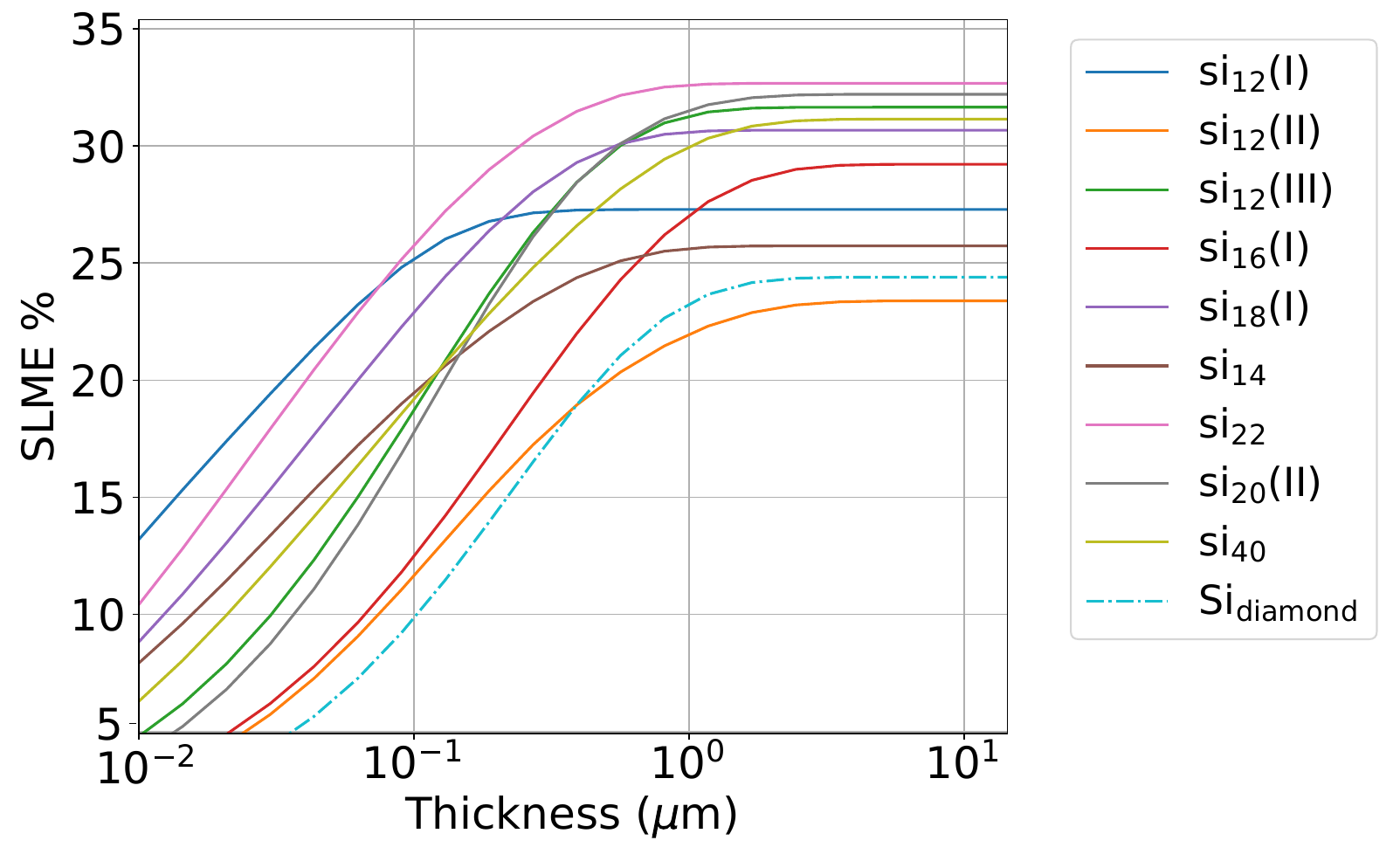}
	\caption{The spectroscopic limited maximum efficiency (SLME) as a function of thickness for the most promising structures. The dash-dot line corresponds to Si$_{\text{diamond}}$, for which we set the value of ($f_r$) to the suggested value of ($10^{-3}$). All the calculations were performed at 300K.}
	\label{slme}
\end{figure}

In Figure \ref{slme}, SLME as a function of thickness for the promising structures in this study is displayed. For comparison, the graph for Si$_{\text{diamond}}$ is also shown. As evident from the figure, all structures except Si$_{12}$(II) exhibit higher efficiency at thicknesses greater than 1$\mu m$ compared to Si$_{\text{diamond}}$. The highest efficiencies are observed for Si$_{22}$, Si$_{20}$(II), Si$_{12}$(III), and Si$_{40}$, respectively. Although Si$_{12}$(I) shows higher efficiency initially, its efficiency decreases relative to other structures as thickness increases. These calculations were performed at 300K.

\begin{figure}
	\centering%
	\includegraphics[scale=.38]{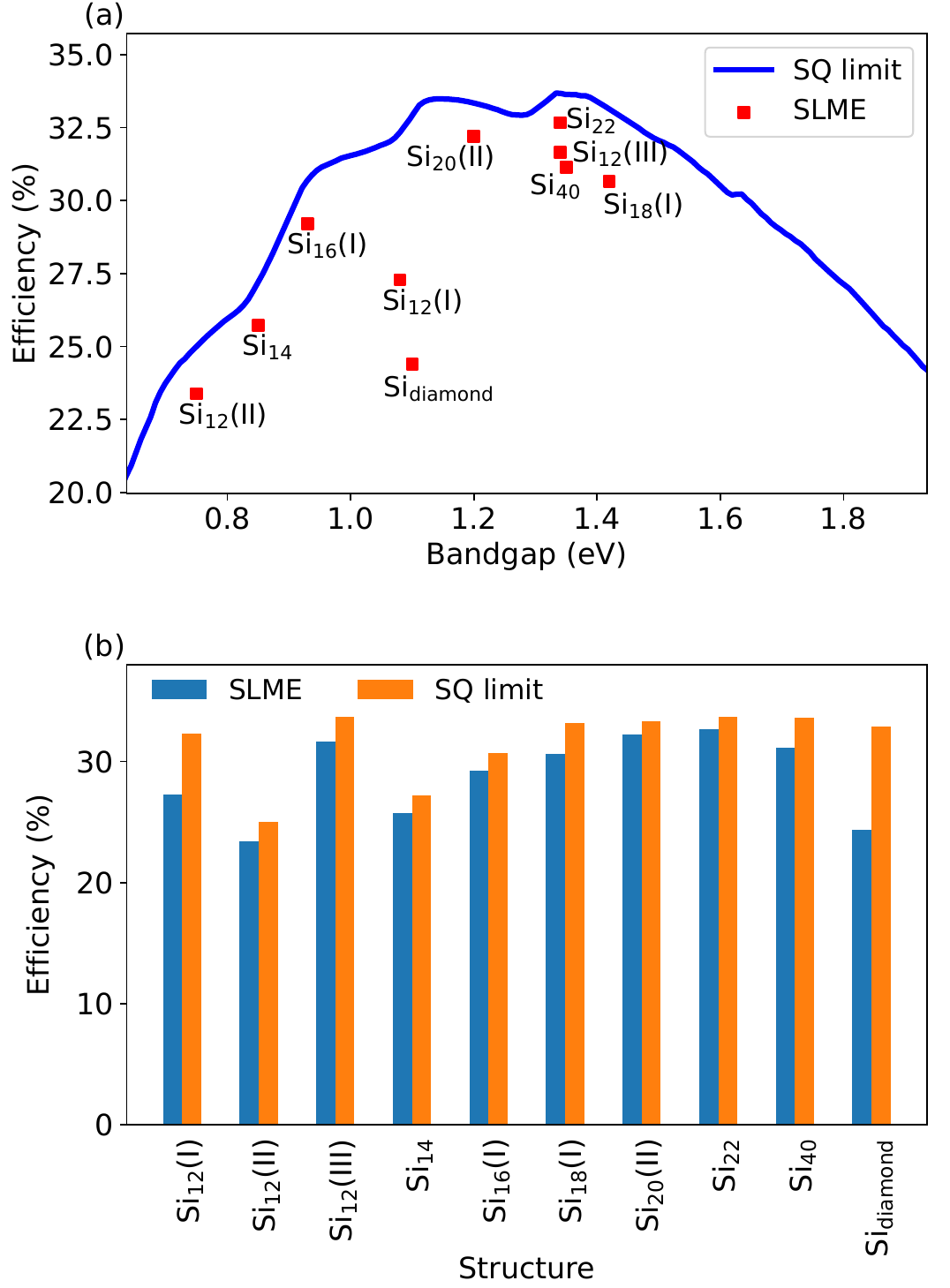}
	\caption{(a) Shockley-Queisser (SQ) limit diagram as a function of the band gap value. The red dots represent the SLME values for the structures. (b) Bar plot showing the efficiency of structures according to the SQ limit (orange bars) and SLME (blue bars). All calculations were performed at 300 K, with the device thickness kept constant at 500 nm.}
	\label{sq-slme}
\end{figure}

Figure \ref{sq-slme}(a) shows the Shockley-Queisser (SQ) limit and SLME values for the structures. 
In Figure \ref{sq-slme}(b), the bar plot of SLME and efficiency values corresponding to the SQ limit for the structures is displayed. The SLME values of the structures were measured at a temperature of 300 K and a thickness of 500 nm. As evident, structures Si$_{22}$, Si$_{20}$(II), Si$_{12}$(III), Si$_{40}$, and Si$_{18}$(I) have the highest efficiencies, with SLME values of 32.67\%, 32.20\%, 31.64\%, 31.14\%, and 30.66\%, respectively.


\subsection{Candidate Structures}
Si$_{12}$(III) stands out as the top choice for a solar cell when considering both stability and GW band gap. 
However, if we focus solely on the GW band gap, Si$_{12}$(III), Si$_{18}$(I), Si$_{18}$(II), Si$_{22}$, 
Si$_{36}$(I), and Si$_{40}$ are potential candidates. 
Among these, Si$_{40}$ appears promising for three reasons:
I) Various methods indicate a band gap of $\sim1.35$ eV for Si$_{40}$.
II) The structure is significantly different from Si$_{\text{diamond}}$ in the Born-Oppenheimer energy landscape, 
making the transformation to Si$_{\text{diamond}}$ challenging due to a large barrier.
III) Si$_{40}$ is ductile, making it practical for flexible solar cells. 
%In addition, we can highlight the most promising structures based on the absorption spectrum of Si$_{12}$(I) and Si$_{14}$. Although, according to SLME values, Si$_{22}$, Si$_{20}$(II), Si$_{12}$(III), Si$_{40}$, and Si$_{18}$(I) are the most suitable structures for solar cell production, respectively, because their SLME  is more than 30\%.
Additionally, based on the absorption spectrum, we can identify Si$_{12}$(I) and Si$_{14}$ as the most promising structures. 
However, based on SLME values, the structures Si$_{22}$, Si$_{20}$(II), Si$_{12}$(III), Si$_{40}$, and Si$_{18}$(I) are the most suitable for solar cell production, as their SLME is above 30\%.

\subsection{Porous Structures with Void Tunnels}

\begin{figure*}[htbp]
	\centering%
    \includegraphics[scale=.1]{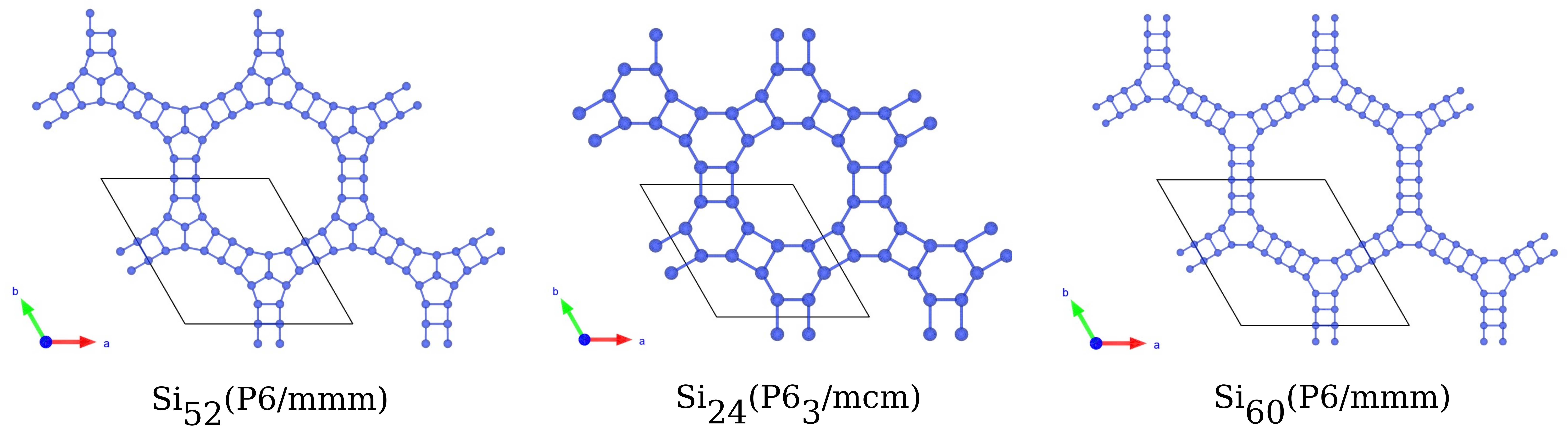}
	\caption{The ball-and-stick diagrams of the crystal structures.}
	\label{tunnel}
\end{figure*}
In our research, we discover specific porous structures, like Si$_{40}$, characterized by void tunnels, 
setting them apart from others. These structures exhibit hexagonal symmetry. 
To explore similar structures within this symmetry, we employ an evolutionary algorithm.
Our investigation leads us to identify two stable structures (Si$_{60}$ and Si$_{24}$). 
Additionally, we attempted to create Si$_{52}$ by expanding upon the Si$_{40}$ tunnel.
These structures are illustrated in Figure~\ref{tunnel}. 
Analyzing their properties, we found the HSE band gaps for Si$_{24}$, 
Si$_{52}$, and Si$_{60}$ to be 1.57, 1.00, and 0.76 eV, respectively.

These outcomes demonstrate a pattern: as the tunnel radius increases, the band gap decreases. 
This phenomenon aligns with the concept of quantum confinement.
Thus, manipulating the tunnel radius offers a means to engineer and adjust the band gap.

\subsection{Machine Learning Calculations}
The initial step in employing ML for predicting material properties involves the preparation of input data. The generation of a suitable input derived from a material has been a focal point of discussion among researchers. 
In general, there are two basic approaches to construct the feature vector (or descriptor) of a material. The first approach uses compositional information, typically derived from the elemental and chemical properties and the stoichiometry of the structure\cite{Zhang2018,10.1021/acs.chemmater.7b00789,Jha2018,PhysRevB.98.214112}. The second approach relies on structural information, such as atomic positions, exemplified by the Coulomb matrix\cite{PhysRevLett.108.058301} and MBTR descriptors.
However, since in this study, all the data contain the same element (Si), the first approach is not applicable. This limitation provides an excellent opportunity to assess the effectiveness of the second approach. Additionally, we have generated descriptor for each structure from a set of global structural properties, such as volume and energy of the structures. This was accomplished by extracting 24 features from the outputs of DFT calculations.

Furthermore, structural descriptors like MBTR and SOAP can be broadly categorized into two main types: global and local descriptors~\cite{Damewood2023, HIMANEN2020106949}. 
When predicting a property of a material, it's crucial to consider whether the property is global or local in nature. 
This is important because attempting to predict a global property using only local descriptors can lack physical meaning. 
While it might be feasible mathematically, it may not make sense from a physical perspective. 
If we intend to use a local descriptor to predict a global property, such as band gap, we need to find a way to transform the descriptor into a global one.  
This transformation is often achieved by performing operations like inner or outer products on the feature vectors of different local neighborhoods within the structure.
As it detailed in the literatures, the MBTR serves as a global descriptor, while the SOAP is a local descriptor\cite{Damewood2023, HIMANEN2020106949}. 
Consequently, in this study, we have converted the SOAP representation into a global representation to make it suitable for predicting global properties like band gap. 

In addition to these geometric descriptors, in this work, we use other features obtained from the results of DFT calculations.
The selected features include the total energy (E), the total all-electron (AE) energy, internal energy, one electron contribution, XC contribution, 
Ewald contribution, Hartree (H) contribution,one-center PAW contribution, PAW Hartree energy of AE, PAW Hartree energy pseudopotential (PS), 
PAW exchange-correlation (xc) energy of AE, PAW xc energy of PS, total E-H with PAW,  total E-XC with PAW, Fermi energy, scf gap, the energy of the first three conduction bands at gamma point denoted as ($\Gamma \_ E^{\prime}_{i}$) and the energy of the first four valance bands at gamma point denoted as ($\Gamma \_ E_{i}$), along with volume of the system.

Table \ref{MLinformation} displays the Root Mean Squared Error (RMSE) values for band gap predictions on both test and training data. All models are obtained from training on 80\% of the structures (156 structures). 
\begin{table*}[h]
	\centering
	\caption{\small The table presents Root Mean Squared Error (RMSE) and R$^2$ values for predictions on both test and training data. These models were trained using data derived from SOAP and MBTR, and features extracted from the output of DFT calculations(DFT output). 80\% of structures are randomly selected for training the models.}
	%\resizebox{\textwidth}{!}{
	%\tiny
	\small
	\begin{tabular}{|c|*{5}{c|}}
		\hline
		descriptor&ML algorithm& RMSE(train)eV&RMSE(test)eV&R$^2$(train)& R$^2$(test)\\\hline
		\multirow{7}{*}{SOAP}&KRR&0.154&0.397&0.91&0.49\\
		&SVR&0.039&0.353&0.91&0.49\\
		&NN&0.183&0.393&0.87&0.45\\
		&CNN&0.341&0.338&0.56&0.49\\
		&DT&0.167&0.593&0.90&0.01\\
		&RF&0.171&0.415&0.89&0.23\\
		&XGBoost&0.113&0.403&0.95&0.27\\\hline
		\multirow{7}{*}{MBTR}&KRR&0.105&0.403&0.96&0.25\\
		&SVR&0.303&0.392&0.65&0.44\\
		&NN&0.207&0.422&0.84&0.29\\
		&CNN&0.285&0.299&0.69&0.60\\
		&DT&0.099&0.623&0.96&0.01\\
		&RF&0.136&0.343&0.93&0.59\\
		&XGBoost&0.009&0.338&0.99&0.49\\\hline
		\multirow{7}{*}{DFT output}&KRR&0.144&0.171&0.93&0.91\\
		&SVR&0.105&0.169&0.96&0.90\\
		&NN&0.050&0.215&0.99&0.81\\
		&CNN&0.091&0.168&0.97&0.90\\
		&DT&0.045&0.195&0.99&0.88\\
		&RF&0.068&0.164&0.98&0.93\\
		&XGBoost&0.011&0.155&0.99&0.91\\
		\hline
	\end{tabular}%}
	\label{MLinformation}
\end{table*}

As evident from the table, models trained on the features derived from DFT calculations exhibit lower RMSE in band gap predictions.  In the context of this notation, "DFT output" signifies the features obtained through DFT calculations. The R$^2$(coefficient of determination) parameter values in the table also indicate the superior performance of these models compared to those trained on data obtained from MBTR and SOAP.
The optimal models derived from SOAP include the KRR and SVR, with RMSE values of 0.397 eV and 0.353 eV, respectively, for the test data. Both models demonstrate R$^2$ parameter values of 0.91 for training data and 0.49 for test data. In the case of MBTR, the RF and XGBoost models emerge as the best performers, with RMSE of 0.343 eV and 0.348 eV, respectively, for the test data. The R$^2$ parameter values for these models are 0.59 and 0.49, respectively, for the test data.
When utilizing DFT output features, the model accuracy significantly improves. The RMSE for all models drops below 0.21 eV for test data, and the R$^2$ parameter for test data exceeds 0.80 across all models.

\begin{figure*}[h]
	\centering
	\includegraphics[width=1.09\textwidth]{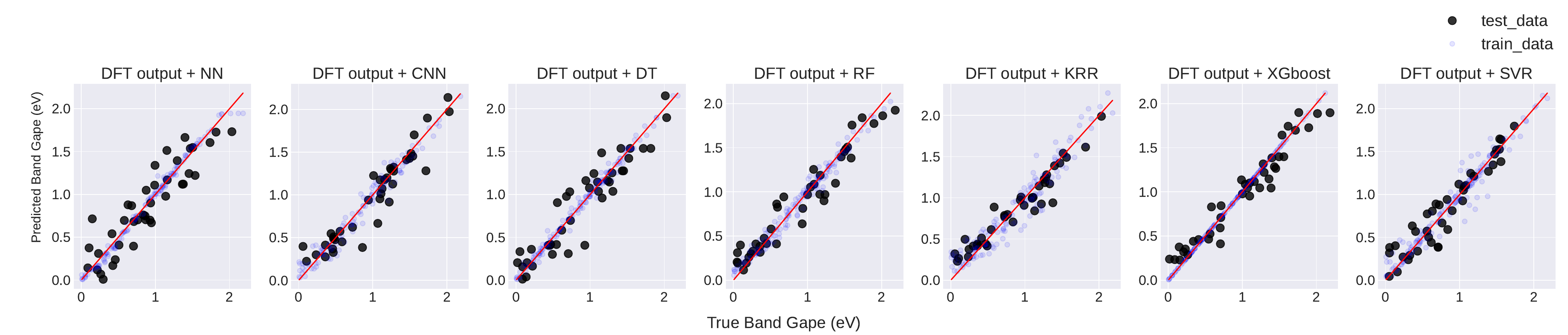}
	\caption{Band gap regression diagram using DFT output features and NN, SVR, CNN, KRR RF, DT, and XGBoost ML models.}\label{ML output}
\end{figure*}

Figure~\ref{ML output} illustrates the relationship between band gap values predicted by ML models (y-axis) and those obtained through DFTB calculations (x-axis). Only models utilizing DFT output features are presented in this figure, while graphs pertaining to models obtained from MBTR and SOAP are included in the SI.

As evident from Figure~\ref{ML output}, the models denoted as "DFT output + XGboost", "DFT output + RF", and "DFT output + CNN" have emerged as the most effective predictors. 

Recognizing the significance of individual features within the context of ML models enhances our ability to interpret the model's performance and its analytical outcomes. 
In this regard, the assessment of models utilizing features extracted from DFT outputs, through the computation of Shapely values\cite{NIPS2017_8a20a862} and the subsequent generation of beeswarm plots, serves as a valuable analytical approach. 
Each model is accompanied by a beeswarm plot, which effectively communicates the relative importance of each feature in the prediction process.
The graphical plots of Shapley values are available in SI.

The examination of these graphical representations unequivocally highlights the feature labeled as "scf\_gap" as possessing a superior degree of importance when compared to the other features under consideration. This insight underscores the pivotal role played by "scf\_gap" in influencing the model's predictive performance.

Upon considering the second most crucial feature for band gap prediction, no consistent ranking was observed across the various models. Specifically, in the KRR and RF models, the "one\_center\_paw" feature claimed the second position, whereas in the XGBoost and DT models, it occupied the third position. Interestingly, in the NN, SVR, and CNN models, "one\_center\_paw" does not appear among the top five important features.

A closer examination of feature importance from the third rank onward reveals a distinct pattern in each model. There appears to be no predetermined order, suggesting that different ML models do not uniformly prioritize features in terms of their influence.

\section{Conclusion}
In this research, our primary objective was to predict and evaluate crystalline silicon semiconductors, 
with a particular focus on their band gap energy properties, with the aim of determining their suitability 
for integration with solar cell technologies, thereby increasing the efficiency of solar energy conversion systems.
After extensive structural searches, we screened the structure according to the band gap using fast DFTB calculations.
%new
We recalculate band gaps using HSE and GW methods. The findings show that the DFTB method with si-band SK is effective for screening band gaps and can be applied in future studies.
%new
To ensure the uniqueness of these newly discovered materials, a comprehensive analysis of their geometric configurations was carefully performed. In addition, we performed a comprehensive evaluation of the mechanical and dynamic stability of these structures. 
Our extensive analyzes yielded compelling evidence of the dynamic stability of these silicon structures, reinforcing their viability for integration into solar cell technologies.
%new
We choose promising candidates, such as Si$_{40}$ and Si$_{12}$, 
with potential applications in solar cells. 
Specifically, structures like Si$_{40}$ show void tunnels, 
suggesting not only an optimal band gap but also structural flexibility.
%new
At the end, We used geometric properties and DFT-derived properties to model and predict band gap energies. 
Subsequently, we performed a critical assessment of the importance of selected features obtained from the DFT outputs, and identified those elements that have the greatest impact on the band gap prediction.
In conclusion, our study provides a comprehensive exploration of new direct and quasi-direct band gap silicon-based semiconductor structures and positions them as promising candidates for next-generation solar cell technology.

%\Urlmuskip=0mu plus 1mu\relax
\bibliographystyle{apsrev4-1}
\bibliography{refrence.bib} %such as MyReferences
%\newpage
%\subsection{Tables}
%\newif\ifshowtables
\end{document}